# BAYESIAN VARIABLE SELECTION USING COST-ADJUSTED BIC, WITH APPLICATION TO COST-EFFECTIVE MEASUREMENT OF QUALITY OF HEALTH CARE

By D. Fouskakis, I. Ntzoufras and D. Draper

*National Technical University of Athens, Athens University of Economics and Business and University of California*

In the field of quality of health care measurement, one approach to assessing patient sickness at admission involves a logistic regression of mortality within 30 days of admission on a fairly large number of sickness indicators (on the order of 100) to construct a sickness scale, employing classical variable selection methods to find an "optimal" subset of 10–20 indicators. Such "benefit-only" methods ignore the considerable differences among the sickness indicators in cost of data collection, an issue that is crucial when admission sickness is used to drive programs (now implemented or under consideration in several countries, including the U.S. and U.K.) that attempt to identify substandard hospitals by comparing observed and expected mortality rates (given admission sickness). When both data-collection cost and accuracy of prediction of 30-day mortality are considered, a large variable-selection problem arises in which costly variables that do not predict well enough should be omitted from the final scale.

In this paper (a) we develop a method for solving this problem based on posterior model odds, arising from a prior distribution that (1) accounts for the cost of each variable and (2) results in a set of posterior model probabilities that corresponds to a generalized cost-adjusted version of the Bayesian information criterion (BIC), and (b) we compare this method with a decision-theoretic cost-benefit approach based on maximizing expected utility. We use reversible-jump Markov chain Monte Carlo (RJMCMC) methods to search the model space, and we check the stability of our findings with two variants of the MCMC model composition ($MC^3$) algorithm. We find substantial agreement between the decision-theoretic and cost-adjusted-BIC methods; the latter provides a principled approach to performing a

> cost-benefit trade-off that avoids ambiguities in identification of an
> appropriate utility structure. Our cost-benefit approach results in a
> set of models with a noticeable reduction in cost and dimensionality,
> and only a minor decrease in predictive performance, when compared
> with models arising from benefit-only analyses.

**1. Introduction.** An important topic in health policy is the assessment of the quality of health care offered to hospitalized patients. Quality of care is usually thought to depend mainly on three ingredients [e.g., Donabedian and Bashshur, (2002)]: (i) *process*, which is what health care providers do on behalf of patients, (ii) *outcomes*, which are what happens to patients as a result of the care they receive, and (iii) patient *sickness at admission*, because the appropriateness of outcomes cannot be judged without taking account of the burden of illness brought to the hospital by its patients.

A direct audit of the processes of care is usually regarded as the single most informative component in an evaluation of quality, but process is much more expensive to measure than outcomes or admission sickness [e.g., Kahn et al. (1990a); Schuster et al. (2005)]. Interest has therefore focused in recent years, in countries such as the United States and the United Kingdom, on an indirect method of assessment—which might be termed the *input-output* approach[1] [e.g., Draper (1995); Olhssen et al. (2007)]—in which hospital outcomes (for instance, death within 30 days of admission) are compared after adjusting for differences in inputs (sickness at admission). The idea is to treat what goes on inside the hospital—process—as a black box, with the contents of the box inferred by examining its outputs after taking account of its inputs.

1.1. *Indirect measurement of quality of health care.* In practice, to indirectly measure quality of care at any given moment in time, this strategy proceeds by (a) taking a sample of hospitals and a sample of patients in the chosen hospitals, (b) obtaining mortality outcomes for the sampled patients (e.g., from central government data bases), (c) extracting information on admission sickness from the medical records of these patients, (d) forming an expected mortality rate for each hospital based on (c), and (e) comparing observed and expected mortality rates to identify unusual hospitals (on both the "good" and "bad" ends of the spectrum). Since this would involve abstracting data from the charts of many thousands of patients if it were attempted on a large scale, the *cost-effective* measurement of admission sickness is crucial to this approach. Progress is being made in the U.S. [see, e.g.,

---

[1] In the U.K. this approach is also referred to as *league-table quality assessment* [e.g., Goldstein and Spiegelhalter (1996)], by analogy with the process of ranking football (soccer) teams; in the U.S. and elsewhere it is also called *provider profiling* [e.g., Zhang et al. (2006)].

CMS (2008), for details on Medicare's plans to compile a Uniform Clinical Data Set] and elsewhere on real-time electronic data collection of clinically richer sets of process and sickness variables for hospital patients than those previously available from administrative data bases, but it is likely to remain true for at least the next decade that cost-effective collection of data from nonelectronic medical records will be relevant to the design of quality of care studies in health policy [see, e.g., NDNQI (2008), and CalNOC (2008), for current examples, in the field of nursing quality assessment, where extensive nonautomated primary data collection is both ongoing and planned]. This is particularly true in countries with an interest in quality of care measurement but insufficient resources to be at the cutting edge in medical informatics.

Quality of care assessment is a highly disease-specific activity: for instance, the best admission sickness variables to examine for pneumonia would be quite different from those for heart attack.[2] With any given disease there will be on the order of 100 separate variables potentially available in the medical record that are directly or indirectly related to admission sickness. In the case of pneumonia, for example, on which we focus in this paper, a list of the important variables from a clinical perspective would include such things as systolic blood pressure on day 1 of admission, the presence or absence of shortness of breath, and the blood urea nitrogen level (a measure of kidney functioning).

1.2. *Standard benefit-only variable-selection approach.* One standard method for creating an expected mortality rate from these admission sickness inputs is logistic regression, with 30-day death as the outcome, and using a nationally-representative sample of patients to normalize the expectation to average care across the nation. Typically a frequentist variable-selection method—such as backward selection from the model with all predictors—is employed to find a parsimonious and clinically reasonable subset of the available sickness variables. In a major U.S. study conducted by the RAND Corporation, of quality of hospital care for 16,758 elderly patients in the late 1980s [Kahn et al. (1990b)], this approach was used to reduce the initial list of $p = 83$ available sickness indicators gathered on the $n = 2{,}532$ pneumonia patients in the study down to a core of 14 predictors [Keeler et al. (1990)].

As good as the resulting scale may be on grounds of simplicity and ease of clinical communication, we take the view in this paper that—when the goal is the creation of a sickness scale that may be used prospectively to measure quality of care on a new set of patients not yet examined—the original RAND approach is sub-optimal, because it takes no account of differences in the *cost of data collection* among the available predictors (which varied for pneumonia from 30 seconds to 10 minutes of abstraction time per variable). The RAND approach represents a kind of benefit-only analysis; we propose a cost-benefit analysis, in which variables are chosen for the final scale only when they predict mortality well enough given how much they cost to collect. The relevance of this cost-benefit perspective is seen by noting that, in practice, the amount of money devoted to quality assessment will almost invariably be constrained, so that money wasted on excess data collection costs could be better spent on obtaining (e.g.) a larger sample size at the patient and/or hospital levels.

---

[2] Note that this approach to quality assessment is effective only with diseases, such as pneumonia and heart attack, for which there is a strong process-outcome link (i.e., such that good care leads to good outcomes and bad care to bad outcomes); with an incurable illness given present medical understanding (such as end-stage renal disease), mortality is irrelevant, since all patients die relatively quickly no matter what processes of care they receive.

Table 1 lists the 14 variables chosen by the benefit-only RAND approach, together with their marginal data collection costs per patient (expressed in minutes of data abstraction time; this could be transformed to a monetary scale with a map of the form $c \mapsto \alpha c$ with $\alpha > 0$, using the prevailing wage rate for qualified data abstraction personnel, but there is nothing to be gained from such a transformation). The full list of all 83 sickness indicators for pneumonia is examined in Section 4.2, where columns 4–6 of Table 1 are explained.

1.3. *Cost-benefit and cost-restriction-benefit analyses.* Weighing data-collection costs against the accuracy of prediction creates a large variable-selection problem; for example, with $p = 83$ it is necessary to compare $2^{83} \doteq 9.7 \cdot 10^{24}$ subsets of sickness variables in order to find the optimal subset. Solving this problem by brute-force examination of all $10^{25}$ models is sharply infeasible given contemporary computing resources.

Following Fouskakis (2001), suppose (a) the 30-day mortality outcome $Y_i$ and data on $p$ sickness indicators $(X_{i1}, \ldots, X_{ip})$ have been collected on $n$ individuals chosen exchangeably from a population $\mathcal{P}$ of patients with a given disease, and (b) the goal is to predict the death outcome for $n^*$ new patients who will in the future be sampled randomly from $\mathcal{P}$, (c) on the basis of some or all of the predictors $X_j$, when (d) the marginal costs of data collection per patient $c_1, \ldots, c_p$ for the $X_j$ vary considerably. What is the best subset of the $X_j$ to choose based on both the quality and the cost of obtaining the predictions?

In problems such as this, in which there are two desirable criteria—in this case, low cost and high predictive accuracy—that compete, and over which a joint optimization must be achieved, there are two main ways to proceed:

(a) (cost-benefit) both criteria can be placed on a common scale, trading one off against the other, and optimization can occur on that scale, or

TABLE 1
*The RAND admission sickness scale for pneumonia ($p = 14$ variables), with the marginal data collection costs per patient for each variable (in minutes of abstraction time); columns 4–6 are explained in Section 4.2*

| | | | | Method | |
|---|---|---|---|---|---|
| | **Variable** | | Utility | **RJMCMC** | |
| **Index** | **Name** | Cost (minutes) | Good? | Good? | Posterior probability |
| 1 | Systolic blood pressure score (2-point scale) | 0.5 | ** | ** | 0.99 |
| 2 | Age | 0.5 | * | ** | 0.99 |
| 3 | Blood urea nitrogen | 1.5 | ** | ** | 1.00 |
| 4 | APACHE II coma score (3-point scale) | 2.5 | ** | ** | 1.00 |
| 5 | Shortness of breath day 1 (yes, no) | 1.0 | ** | ** | 0.99 |
| 6 | Serum albumin score (3-point scale) | 1.5 | * | ** | 0.55 |
| 7 | Respiratory distress (yes, no) | 1.0 | * | ** | 0.92 |
| 8 | Septic complications (yes, no) | 3.0 | | | 0.00 |
| 9 | Prior respiratory failure (yes, no) | 2.0 | | | 0.00 |
| 10 | Recently hospitalized (yes, no) | 2.0 | | | 0.00 |
| 12 | Initial temperature | 0.5 | * | ** | 0.95 |
| 17 | Chest X-ray congestive heart failure score (3-point scale) | 2.5 | | | 0.00 |
| 18 | Ambulatory score (3-point scale) | 2.5 | | | 0.00 |
| 48 | Total APACHE II score (36-point scale) | 10.0 | | | 0.00 |

(b) (cost-restriction-benefit) one criterion can be optimized, subject to a bound on the other.

Here we present results from one possible implementation of approach (a), in which we identify a prior distribution that (1) accounts for the cost of each variable in a natural way and (2) results in a set of posterior model probabilities that correspond to a generalized cost-adjusted version of the Bayesian information criterion (BIC). To incorporate preferences based on costs of the variables, we use a Laplace approximation to obtain a cost-based penalty for each variable. After setting up the prior model and variable probabilities, we use reversible-jump Markov chain Monte Carlo to search the model space. The data on which we demonstrate our method in this paper consist of the representative sample of $n = 2{,}532$ elderly American patients hospitalized in the period 1980–86 with pneumonia taken from the RAND study described above.

The plan of the paper is as follows. In Section 2 we describe the approach we investigated in this study, and Section 3 provides details concerning the computation. Section 4 illustrates the experimental results on the pneumonia data set using the method described in Sections 2 and 3, and includes a comparison of the results from our method and those from another possible implementation [Fouskakis and Draper (2008)] of approach (a) based on maximizing expected utility. [Brown et al. (1998) presented an application of decision theory to variable selection in multivariate regression that is motivated by somewhat similar cost-benefit considerations in a quite different setting; Lindley (1968) used squared-error loss to measure predictive accuracy while recommending a cost-benefit tradeoff in variable selection in a less problem-specific framework than the one presented here.] In Section 5 we conclude the paper with a brief discussion of some statistical and quality assessment implications of our work.

**2. A Bayesian approach to cost-effective variable selection.** Bayesian parametric model comparison and variable selection are based on specifying a model $m$, its likelihood $f(\mathbf{y}|\boldsymbol{\theta}_m, m)$, the prior distribution of model parameters $f(\boldsymbol{\theta}_m|m)$ and the corresponding prior model weight (or probability) $f(m)$, where $\boldsymbol{\theta}_m$ is a parameter vector under model $m$ and $\mathbf{y}$ is the data vector. Parametric inference is based on the posterior distribution $f(\boldsymbol{\theta}_m|\mathbf{y}, m)$, and quantifying model uncertainty by estimating the posterior model probabilities $f(m|\mathbf{y})$ is also an important issue. Hence, when we consider a set of competing models $\mathcal{M} = \{m_1, m_2, \ldots, m_{|\mathcal{M}|}\}$, we focus on the posterior probability of model $m \in \mathcal{M}$, defined as

$$
(1) \quad f(m|\mathbf{y}) = \frac{f(\mathbf{y}|m)f(m)}{\sum_{m_\ell \in \mathcal{M}} f(\mathbf{y}|m_\ell)f(m_\ell)} = \left( \sum_{m_\ell \in \mathcal{M}} PO_{m_\ell, m} \right)^{-1}
$$
$$
= \left[ \sum_{m_\ell \in \mathcal{M}} B_{m_\ell, m} \frac{f(m_\ell)}{f(m)} \right]^{-1},
$$

where $PO_{m_k, m_\ell} = \frac{f(m_k|\mathbf{y})}{f(m_\ell|\mathbf{y})}$ is the posterior model odds, and $B_{m_k, m_\ell}$ is the Bayes factor, for comparing models $m_k$ and $m_\ell$. When we limit ourselves in the comparison of only two models we typically focus on $PO_{m_k, m_\ell}$ and $B_{m_k, m_\ell}$, which have the desirable property of insensitivity to the selection of the model space $\mathcal{M}$. By definition, the Bayes factor is the ratio of the posterior model odds over the prior model odds; thus, large values of $B_{m_k, m_\ell}$ (usually greater than 12, say) indicate strong posterior support of model $m_k$ against model $m_\ell$ [for details see, e.g., Raftery (1996)]. The posterior model probabilities and integrated likelihoods $f(\mathbf{y}|m_\ell)$ in (1) are rarely analytically tractable; we use a combination of Laplace approximations [e.g., Bernardo and Smith (1994)] and Markov Chain Monte Carlo (MCMC) methodology [e.g., Green (1995); Han and Carlin (2001); Chipman et al. (2001); Dellaportas et al. (2002); Lopes (2002)] to approximate posterior odds and Bayes factors.

In the problem described in Section 1, we use a simple logistic regression model with response $Y_i = 1$ if patient $i$ dies within 30 days of admission and 0 otherwise. We further denote by $X_{ij}$ the sickness predictor variable $j$ for patient $i$ and by $\gamma_j$ an indicator, often used in Bayesian variable selection problems [e.g., George and McCulloch (1993); Kuo and Mallick (1998); Brown et al. (1998); Dellaportas et al. (2002)], taking the value 1 if variable $j$ is included in the model and 0 otherwise. Thus, in this case $\mathcal{M} = \{0,1\}^p$, where $p$ is the total number of variables. In order to map the set of binary model indicators $\boldsymbol{\gamma}$ onto a model $m$, we can use a representation of the form $m(\boldsymbol{\gamma}) = \sum_{j=1}^{p} 2^{j-1} \gamma_j$. Hence, the model formulation can be summarized as

$$
(Y_i | \boldsymbol{\gamma}) \stackrel{\text{indep}}{\sim} \text{Bernoulli}[p_i(\boldsymbol{\gamma})],
$$

(2)
$$
\eta_i(\boldsymbol{\gamma}) = \log\left[\frac{p_i(\boldsymbol{\gamma})}{1 - p_i(\boldsymbol{\gamma})}\right] = \sum_{j=0}^{p} \beta_j \gamma_j X_{ij},
$$

$$
\eta(\boldsymbol{\gamma}) = \mathbf{X} \operatorname{diag}(\boldsymbol{\gamma}) \beta = \mathbf{X}_{\boldsymbol{\gamma}} \boldsymbol{\beta}_{\boldsymbol{\gamma}},
$$

defining $X_{i0} = 1$ for all $i = 1, \ldots, n$ and $\gamma_0 = 1$ with prior probability one since here the intercept is always included in all models. Here $p_i(\boldsymbol{\gamma})$ is the death probability (which may be thought of as the sickness score) for patient $i$ under model $\boldsymbol{\gamma}$, $\eta(\boldsymbol{\gamma}) = [\eta_1(\boldsymbol{\gamma}), \ldots, \eta_n(\boldsymbol{\gamma})]^T$, $\boldsymbol{\gamma} = (\gamma_0, \gamma_1, \ldots, \gamma_p)^T$, $\boldsymbol{\beta} = (\beta_0, \beta_1, \ldots, \beta_p)^T$, and $\mathbf{X} = (X_{ij}, i = 1, \ldots, n; j = 0, 1, \ldots, p)$. The vector $\boldsymbol{\beta}_{\boldsymbol{\gamma}}$ stands for the subvector of $\beta$ included in the model specified by $\boldsymbol{\gamma}$, that is, $\boldsymbol{\beta}_{\boldsymbol{\gamma}} = (\beta_j : \gamma_j = 1, j = 0, 1, \ldots, p)$, and is equivalent to the $\boldsymbol{\theta}_m$ vector defined at the beginning of this section; similarly, $\mathbf{X}_{\boldsymbol{\gamma}}$ is the submatrix of $\mathbf{X}$ with columns corresponding to variables included in the model specified by $\boldsymbol{\gamma}$.

In the remainder of this section we illustrate how to build a prior distribution to accommodate in the posterior distribution a penalty function for the increased cost of expensive predictor variables. To this end, we first build a minimally informative prior for the model parameters based on the ideas of Ntzoufras et al. (2003). Then we employ a Laplace approximation [e.g., Tierney and Kadane (1986)] to examine the penalty (indirectly) imposed upon the model likelihood using the Bayesian approach. Finally, we specify prior model weights (probabilities) in such a way that the posterior model probabilities in effect result from a likelihood penalized according to the cost of each variable in the model.

2.1. *Prior on model parameters.* One important problem in Bayesian model evaluation using posterior model probabilities is their sensitivity to the prior variance of the model parameters: large variance of the $\boldsymbol{\beta_\gamma}$ (used to represent prior ignorance) will increase the posterior probabilities of the simpler models considered in the model space $\mathcal{M}$ [Bartlett (1957); Lindley (1957); Shafer (1982); Robert (1993); Kass and Raftery (1995); Sinharay and Stern (2002)]. Therefore, specifying the prior distribution is pivotal for the a posteriori support of the models examined. We address this issue with ideas proposed by Ntzoufras et al. (2003): we use a prior distribution of the form

$$(3) \qquad f(\boldsymbol{\beta_\gamma}|\boldsymbol{\gamma}) = N(\mu_{\boldsymbol{\gamma}}, \boldsymbol{\Sigma_\gamma}),$$

with prior covariance matrix given by $\boldsymbol{\Sigma_\gamma} = n[\mathcal{I}(\boldsymbol{\beta_\gamma})]^{-1}$, where $n$ is the total sample size and $\mathcal{I}(\boldsymbol{\beta_\gamma})$ is the information matrix

$$(4) \qquad \mathcal{I}(\boldsymbol{\beta_\gamma}) = \mathbf{X}_{\boldsymbol{\gamma}}^T \mathbf{W}_{\boldsymbol{\gamma}} \mathbf{X}_{\boldsymbol{\gamma}};$$

here $\mathbf{W}_{\boldsymbol{\gamma}}$ is a diagonal matrix, which in the Bernoulli case [e.g., McCullagh and Nelder (1983)] takes the form

$$(5) \qquad \mathbf{W}_{\boldsymbol{\gamma}} = \text{diag}\{p_i(\boldsymbol{\gamma})[1 - p_i(\boldsymbol{\gamma})]\}.$$

This is the *unit information prior* introduced by Kass and Wasserman (1996), which corresponds to adding one data point to the data. Here we use this prior as a base, but we specify $p_i(\boldsymbol{\gamma})$ in the information matrix according to our prior information. In this manner we avoid (even minimal) reuse of the data in the prior.

When little prior information is available, a reasonable prior mean for $\boldsymbol{\beta_\gamma}$ is $\boldsymbol{\mu_\gamma} = \mathbf{0}$. This corresponds to a prior mean on the log-odds scale of zero, from which a sensible prior estimate for all model probabilities is $p_i(\boldsymbol{\gamma}) = 1/2$; with this choice (3) becomes

$$(6) \qquad f(\boldsymbol{\beta_\gamma}|\boldsymbol{\gamma}) = N[\mathbf{0}, 4n(\mathbf{X}_{\boldsymbol{\gamma}}^T \mathbf{X}_{\boldsymbol{\gamma}})^{-1}].$$

This prior distribution can also be motivated by combining the idea of imaginary data with the power-prior approach of Chen et al. (2000); for details, see Fouskakis, Ntzoufras and Draper (2009a).

2.2. *A cost-penalized prior on model space.* The aim of this subsection is to specify a set of prior model probabilities (or odds) that accounts for prior preferences based on the variable costs. To make this more explicit, we first describe preliminary results concerning the posterior model probabilities $f(\boldsymbol{\gamma}|\mathbf{y})$ and the corresponding model odds using the prior distribution (6), when no assumption is made for the prior model probability $f(\boldsymbol{\gamma})$. We then specify a prior on the model space that takes into account prior preferences based on the cost of the variables. In order to achieve this, we use a penalty-based interpretation of the prior $f(\boldsymbol{\gamma})$ imposed on the log-likelihood that directly results from the first subsection. Finally, we use this cost-penalized model prior to calculate the posterior model probabilities and odds.

2.2.1. *Preliminary results: posterior probabilities and model odds in the general setup.* Let us denote by $PO_{k\ell}$ and $B_{k\ell}$ the posterior odds and Bayes factor respectively of model $\boldsymbol{\gamma}^{(k)}$ versus model $\boldsymbol{\gamma}^{(\ell)}$. Then we have

$$
\begin{aligned}
-2\log PO_{k\ell} &= -2[\log f(\boldsymbol{\gamma}^{(k)}|\mathbf{y}) - \log f(\boldsymbol{\gamma}^{(\ell)}|\mathbf{y})] \\
&= -2\log\left(\frac{f(\mathbf{y}|\boldsymbol{\gamma}^{(k)})}{f(\mathbf{y}|\boldsymbol{\gamma}^{(\ell)})}\right) - 2\log\frac{f(\boldsymbol{\gamma}^{(k)})}{f(\boldsymbol{\gamma}^{(\ell)})} \\
&= -2\log B_{k\ell} + \xi(\boldsymbol{\gamma}^{(k)}, \boldsymbol{\gamma}^{(\ell)}),
\end{aligned}
\tag{7}
$$

where $\xi(\boldsymbol{\gamma}^{(k)}, \boldsymbol{\gamma}^{(\ell)})$ is the extra penalty imposed on minus twice the logarithm of the Bayes factor through the prior model probabilities.

Following the approach of Raftery (1996), we can approximate the posterior distribution of a model $\boldsymbol{\gamma}$ using the following Laplace approximation:

$$
\begin{aligned}
-2\log f(\boldsymbol{\gamma}|\mathbf{y}) = &-2\log f(\mathbf{y}|\tilde{\boldsymbol{\beta}}_{\boldsymbol{\gamma}}, \boldsymbol{\gamma}) - 2\log f(\tilde{\boldsymbol{\beta}}_{\boldsymbol{\gamma}}|\boldsymbol{\gamma}) - d_{\boldsymbol{\gamma}}\log(2\pi) \\
&-\log|\boldsymbol{\Psi}_{\boldsymbol{\gamma}}| - 2\log f(\boldsymbol{\gamma}) + O(n^{-1}),
\end{aligned}
\tag{8}
$$

where $\tilde{\boldsymbol{\beta}}_{\boldsymbol{\gamma}}$ is the posterior mode of $f(\boldsymbol{\beta}_{\boldsymbol{\gamma}}|\mathbf{y}, \boldsymbol{\gamma})$, $d_{\boldsymbol{\gamma}} = \sum_{j=0}^{p}\gamma_j$ is the dimension of the model $\boldsymbol{\gamma}$, and $\boldsymbol{\Psi}_{\boldsymbol{\gamma}}$ is minus the inverse of the Hessian matrix of $h(\boldsymbol{\beta}_{\boldsymbol{\gamma}}) = \log f(\mathbf{y}|\boldsymbol{\beta}_{\boldsymbol{\gamma}}, \boldsymbol{\gamma}) + \log f(\boldsymbol{\beta}_{\boldsymbol{\gamma}}|\boldsymbol{\gamma})$ evaluated at the posterior mode $\tilde{\boldsymbol{\beta}}_{\boldsymbol{\gamma}}$. Under the model formulation given by equation (2) and the prior distribution (6), we have that

$$
\begin{aligned}
\boldsymbol{\Psi}_{\boldsymbol{\gamma}} &= \left[-\frac{\partial^2 \log f(\mathbf{y}|\boldsymbol{\beta}_{\boldsymbol{\gamma}}, \boldsymbol{\gamma})}{\partial \boldsymbol{\beta}_{\boldsymbol{\gamma}}^2}\bigg|_{\boldsymbol{\beta}_{\boldsymbol{\gamma}}=\tilde{\boldsymbol{\beta}}_{\boldsymbol{\gamma}}} - \frac{\partial^2 \log f(\boldsymbol{\beta}_{\boldsymbol{\gamma}}|\boldsymbol{\gamma})}{\partial \boldsymbol{\beta}_{\boldsymbol{\gamma}}^2}\bigg|_{\boldsymbol{\beta}_{\boldsymbol{\gamma}}=\tilde{\boldsymbol{\beta}}_{\boldsymbol{\gamma}}}\right]^{-1} \\
&= \left(\mathbf{X}_{\boldsymbol{\gamma}}^T \operatorname{diag}\left\{\frac{\exp(\mathbf{X}_{\boldsymbol{\gamma},i}\tilde{\boldsymbol{\beta}}_{\boldsymbol{\gamma}})}{[1+\exp(\mathbf{X}_{\boldsymbol{\gamma},i}\tilde{\boldsymbol{\beta}}_{\boldsymbol{\gamma}})]^2} + \frac{1}{4n}\right\}\mathbf{X}_{\boldsymbol{\gamma}}\right)^{-1},
\end{aligned}
\tag{9}
$$

where $\mathbf{X}_{\boldsymbol{\gamma},i}$ is row $i$ of the matrix $\mathbf{X}_{\boldsymbol{\gamma}}$ for $i = 1, \ldots, n$.

By substituting the prior (6) in expression (8), we obtain

$$
-2\log f(\boldsymbol{\gamma}|\mathbf{y}) = -2\log f(\mathbf{y}|\tilde{\boldsymbol{\beta}}_{\boldsymbol{\gamma}}, \boldsymbol{\gamma}) + [\phi(\boldsymbol{\gamma}) - 2\log f(\boldsymbol{\gamma})] + O(n^{-1}),
\tag{10}
$$

where

$$
\phi(\boldsymbol{\gamma}) = \frac{1}{4n}\tilde{\boldsymbol{\beta}}_{\boldsymbol{\gamma}}^T \mathbf{X}_{\boldsymbol{\gamma}}^T \mathbf{X}_{\boldsymbol{\gamma}} \tilde{\boldsymbol{\beta}}_{\boldsymbol{\gamma}} + d_{\boldsymbol{\gamma}}\log(4n) + \log\frac{|\boldsymbol{\Psi}_{\boldsymbol{\gamma}}^{-1}|}{|\mathbf{X}_{\boldsymbol{\gamma}}^T \mathbf{X}_{\boldsymbol{\gamma}}|}.
\tag{11}
$$

From the above expression it is clear that the logarithm of a posterior model probability can be regarded as a penalized log-likelihood evaluated at the posterior mode of the model, in which the term $[\phi(\boldsymbol{\gamma}) - 2\log f(\boldsymbol{\gamma})]$ can be interpreted as the penalty imposed upon the log-likelihood. In pairwise model comparisons, we can directly use the posterior model odds (7), which can now be written as

$$
-2\log PO_{k\ell} = -2\log\left\{\frac{f(\mathbf{y}|\tilde{\boldsymbol{\beta}}_{\boldsymbol{\gamma}^{(k)}},\boldsymbol{\gamma}^{(k)})}{f(\mathbf{y}|\tilde{\boldsymbol{\beta}}_{\boldsymbol{\gamma}^{(\ell)}},\boldsymbol{\gamma}^{(\ell)})}\right\}
$$
(12)
$$
+ \left[\phi(\boldsymbol{\gamma}^{(k)}) - \phi(\boldsymbol{\gamma}^{(\ell)}) - 2\log\frac{f(\boldsymbol{\gamma}^{(k)})}{f(\boldsymbol{\gamma}^{(\ell)})}\right] + O(n^{-1}).
$$

Therefore, the comparison of the two models is based on a penalized log-likelihood ratio, where the penalty is now given by

$$
(13) \qquad \psi(\boldsymbol{\gamma}^{(k)},\boldsymbol{\gamma}^{(\ell)}) = \phi(\boldsymbol{\gamma}^{(k)}) - \phi(\boldsymbol{\gamma}^{(\ell)}) - 2\log\frac{f(\boldsymbol{\gamma}^{(k)})}{f(\boldsymbol{\gamma}^{(\ell)})};
$$

for more details see Ntzoufras (1999), Chapter 6.

Each penalty term is divided into two parts: $\phi(\boldsymbol{\gamma})$ and $-2\log f(\boldsymbol{\gamma})$. The first term, $\phi(\boldsymbol{\gamma})$, has its source in the marginal likelihood $f(\mathbf{y}|\boldsymbol{\gamma})$ of model $\boldsymbol{\gamma}$ and can be thought of as a measure of discrepancy between the data and the prior information for the model parameters. The second part comes from the prior model probabilities $f(\boldsymbol{\gamma})$. Indifference on the space of all models, usually expressed by the uniform distribution [i.e., $f(\boldsymbol{\gamma}) \propto 1$], eliminates the second term from the model comparison procedure, since the penalty term in (12) will then be based only on the difference of the first penalty terms $\phi(\boldsymbol{\gamma}^{(k)}) - \phi(\boldsymbol{\gamma}^{(\ell)})$. For this reason the penalty term $\phi(\boldsymbol{\gamma})$ is the imposed penalty that appears in the penalized log-likelihood expression of the Bayes factor $B_{k\ell}$ with a uniform prior on model space.

A simpler but less accurate approximation of $\log PO_{k\ell}$ can be obtained following the arguments of Schwarz (1978):

$$
-2\log PO_{k\ell} = -2\log\left[\frac{f(\mathbf{y}|\hat{\boldsymbol{\beta}}_{\boldsymbol{\gamma}^{(k)}},\boldsymbol{\gamma}^{(k)})}{f(\mathbf{y}|\hat{\boldsymbol{\beta}}_{\boldsymbol{\gamma}^{(\ell)}},\boldsymbol{\gamma}^{(\ell)})}\right](d_{\boldsymbol{\gamma}^{(k)}} - d_{\boldsymbol{\gamma}^{(\ell)}})\log n
$$
(14)
$$
-2\log\frac{f(\boldsymbol{\gamma}^{(k)})}{f(\boldsymbol{\gamma}^{(\ell)})} + O(1)
$$
$$
= BIC_{k\ell} - 2\log\frac{f(\boldsymbol{\gamma}^{(k)})}{f(\boldsymbol{\gamma}^{(\ell)})} + O(1),
$$

where $BIC_{k\ell}$ is the Bayesian Information Criterion [e.g., Kass and Wasserman (1996); Raftery (1995, 1996); Hoeting et al. (1999)] for choosing between models $\boldsymbol{\gamma}^{(k)}$ and $\boldsymbol{\gamma}^{(\ell)}$ and $\hat{\boldsymbol{\beta}}_{\boldsymbol{\gamma}}$ is the vector of maximum likelihood estimates of $\boldsymbol{\beta}_{\boldsymbol{\gamma}}$. Since $BIC_{k\ell}$ is an $O(1)$ approximation, it might diverge from the exact value of the logarithm of the Bayes factor even for large samples. Even so, it has often been shown to provide a reasonable measure of evidence (for finite $n$) and its straightforward calculation has encouraged its widespread use in practice [see Kass and Raftery (1995) for details].

2.2.2. *Accounting for the cost of variables via prior model weights.* Following the previous section and equations (7), (10) and (12), it is clear that an additional penalty $\xi$ can be directly imposed on terms of the form $-2\log B_{kl}$ via the prior model probabilities $f(\boldsymbol{\gamma})$. Here we propose to specify our prior model probabilities via cost-dependent penalties for each variable. We do this by identifying a baseline cost $c_0$ and then specifying the other costs in relation to the baseline in a way that appropriately generalizes BIC. We specify our prior distribution on $\boldsymbol{\gamma}$ to satisfy the following five criteria:

(a) The marginal costs $(c_1, \ldots, c_p)$ should enter into the prior in a manner that is invariant under maps of the form $c \mapsto \alpha c$ with $\alpha > 0$, so that conversion between time and money (see Section 1.2) or between measurements of money on different scales (e.g., dollars and euros) leaves the prior unchanged;
(b) the extra penalty $\xi_1$ for adding a variable $X_j$ with baseline cost $c_0$, above and beyond that in a benefit-only analysis, is zero;
(c) the extra penalty $\xi_2$ for adding a variable $X_j$ with cost $c_j = \kappa c_0$ for some $\kappa > 1$, above and beyond that in a benefit-only analysis, equals the BIC penalty of $(\kappa - 1)$ variables with cost $c_0$;
(d) the extra penalty $\xi_3$ for adding any variable $X_j$, above and beyond that in a benefit-only analysis, is greater or equal to zero; and
(e) if all the variables have the same cost, then the prior must reduce to the uniform prior on $\boldsymbol{\gamma}$.

The first requirement ensures that the prior is invariant with respect to the manner in which cost is measured. The second criterion ensures that the penalty for adding a variable $X_j$ with baseline cost $c_0$ is the same as in the benefit-only analysis. Concerning the third requirement, the proportionality of the extra penalty to the BIC penalty ($\log n$) ensures that the posterior model odds will still have a BIC-like behavior. Moreover, the extra penalty induced by this type of prior will equal the relative difference between the cost of the variable $X_j$ and a variable with cost equal to $c_0$. The fourth requirement ensures that the cost-benefit analysis will support more parsimonious models, in terms of both dimensionality and cost, than the corresponding models supported by the benefit-only analysis under the uniform prior on the model space. Finally, the fifth criterion requires that our prior should reproduce the benefit-only analysis if all costs are equal.

The following theorem, whose proof is given in Appendix A, provides the only prior that meets the above five requirements, and defines the choice of $c_0$.

THEOREM 1. *If a prior distribution $f(\gamma)$ satisfies requirements* (a)–(e) *above, then it must be of the form*

$$f(\gamma_j) \propto \exp\left[-\frac{\gamma_j}{2}\left(\frac{c_j}{c_0} - 1\right)\log n\right] \qquad \text{for } j = 1, \ldots, p, \tag{15}$$

where $c_j$ is the marginal cost per observation for variable $X_j$ and $c_0 = \min\{c_j, j = 1, \ldots, p\}$.

To the above definition of our prior we add the further assumption that the constant term is included in all models by specifying $f(\gamma_0 = 1) = 1$, resulting in

$$(16) \quad -2\log f(\boldsymbol{\gamma}) = \sum_{j=1}^{p} \gamma_j \frac{c_j}{c_0} \log n - d_{\boldsymbol{\gamma}} \log n + 2\sum_{j=1}^{p} \log[1 + n^{-(1-c_j/c_0)/2}].$$

When comparing two models $\boldsymbol{\gamma}^{(k)}$ and $\boldsymbol{\gamma}^{(\ell)}$, the additional penalty imposed on the log-likelihood ratio due to the cost-adjusted prior model probabilities is given by

$$
\begin{aligned}
-2\log\left[\frac{f(\boldsymbol{\gamma}^{(k)})}{f(\boldsymbol{\gamma}^{(\ell)})}\right] &= \sum_{j=1}^{p}(\gamma_j^{(k)} - \gamma_j^{(\ell)})\frac{c_j}{c_0}\log n - (d_{\boldsymbol{\gamma}^{(k)}} - d_{\boldsymbol{\gamma}^{(\ell)}})\log n \\
&= \left[\frac{C_{\boldsymbol{\gamma}^{(k)}} - C_{\boldsymbol{\gamma}^{(\ell)}}}{c_0} - (d_{\boldsymbol{\gamma}^{(k)}} - d_{\boldsymbol{\gamma}^{(\ell)}})\right]\log n,
\end{aligned}
\tag{17}
$$

where $C_{\boldsymbol{\gamma}} = \sum_{j=1}^{p} \gamma_j c_j$ is the total cost of model $\boldsymbol{\gamma}$; thus, two models of the same dimension and cost will have the same prior weight.

Using the prior model odds (17) in the approximate posterior model odds (12), we obtain

$$(18) \quad -2\log PO_{k\ell} = -2\log\left[\frac{f(\mathbf{y}|\tilde{\boldsymbol{\beta}}_{\boldsymbol{\gamma}^{(k)}}, \boldsymbol{\gamma}^{(k)})}{f(\mathbf{y}|\tilde{\boldsymbol{\beta}}_{\boldsymbol{\gamma}^{(\ell)}}, \boldsymbol{\gamma}^{(\ell)})}\right] + \psi(\boldsymbol{\gamma}^{(k)}, \boldsymbol{\gamma}^{(\ell)}) + O(n^{-1}),$$

where the penalty term is given by

$$
\begin{aligned}
\psi(\boldsymbol{\gamma}^{(k)}, \boldsymbol{\gamma}^{(\ell)}) &= \frac{1}{4n}(\tilde{\beta}_{\boldsymbol{\gamma}^{(k)}}^T \mathbf{X}_{\boldsymbol{\gamma}^{(k)}}^T \mathbf{X}_{\boldsymbol{\gamma}^{(k)}} \tilde{\beta}_{\boldsymbol{\gamma}^{(k)}} - \tilde{\beta}_{\boldsymbol{\gamma}^{(\ell)}}^T \mathbf{X}_{\boldsymbol{\gamma}^{(\ell)}}^T \mathbf{X}_{\boldsymbol{\gamma}^{(\ell)}} \tilde{\beta}_{\boldsymbol{\gamma}^{(\ell)}}) \\
&\quad + (d_{\boldsymbol{\gamma}^{(k)}} - d_{\boldsymbol{\gamma}^{(\ell)}})\log(4) \\
&\quad + \log\frac{|\boldsymbol{\Psi}_{\boldsymbol{\gamma}^{(k)}}^{-1}|}{|\mathbf{X}_{\boldsymbol{\gamma}^{(k)}}^T \mathbf{X}_{\boldsymbol{\gamma}^{(k)}}|} - \log\frac{|\boldsymbol{\Psi}_{\boldsymbol{\gamma}^{(\ell)}}^{-1}|}{|\mathbf{X}_{\boldsymbol{\gamma}^{(\ell)}}^T \mathbf{X}_{\boldsymbol{\gamma}^{(\ell)}}|} + \frac{C_{\boldsymbol{\gamma}^{(k)}} - C_{\boldsymbol{\gamma}^{(\ell)}}}{c_0}\log n.
\end{aligned}
\tag{19}
$$

Finally, we consider the BIC-based approximation (14) to the logarithm of the posterior model odds with the prior model odds (17), yielding

$$(20) \quad -2\log PO_{k\ell} = -2\log\left[\frac{f(\mathbf{y}|\hat{\boldsymbol{\beta}}_{\boldsymbol{\gamma}^{(k)}}, \boldsymbol{\gamma}^{(k)})}{f(\mathbf{y}|\hat{\boldsymbol{\beta}}_{\boldsymbol{\gamma}^{(\ell)}}, \boldsymbol{\gamma}^{(\ell)})}\right] + \frac{C_{\boldsymbol{\gamma}^{(k)}} - C_{\boldsymbol{\gamma}^{(\ell)}}}{c_0}\log n + O(1).$$

The penalty term $d_{\boldsymbol{\gamma}}\log n$ of model $\boldsymbol{\gamma}$ used in (14) has been replaced in the above expression by the cost-dependent penalty $c_0^{-1}C_{\boldsymbol{\gamma}}\log n$; ignoring costs is equivalent to taking $c_j = c_0$ for all $j$, yielding $c_0^{-1} C_{\boldsymbol{\gamma}} = d_{\boldsymbol{\gamma}}$, the original BIC expression. Therefore, we may interpret the quantity $\log n$ as the imposed penalty for each variable included in the model $\boldsymbol{\gamma}$ when no costs are considered (or when costs are equal). Moreover, this baseline penalty term is inflated proportionally to the cost ratio $\frac{c_j}{c_0}$ for each variable $X_j$; for example, if the cost of a variable $X_j$ is twice the minimum cost ($c_j = 2c_0$), then the imposed penalty is equivalent to adding two variables with the minimum cost. For all these reasons, (20) can be considered as a cost-adjusted generalization of BIC when prior model probabilities of type (15) are adopted.

To summarize the effect of our prior on the posterior model odds, consider any two models $\boldsymbol{\gamma}^{(k)}$ and $\boldsymbol{\gamma}^{(\ell)}$. From (14) the penalty imposed on the log-likelihood ratio is given by

$$\omega(\boldsymbol{\gamma}^{(k)}, \boldsymbol{\gamma}^{(\ell)}) = (d_{\boldsymbol{\gamma}^{(k)}} - d_{\boldsymbol{\gamma}^{(\ell)}}) \log n - 2 \log \frac{f(\boldsymbol{\gamma}^{(k)})}{f(\boldsymbol{\gamma}^{(\ell)})}$$
$$= (d_{\boldsymbol{\gamma}^{(k)}} - d_{\boldsymbol{\gamma}^{(\ell)}}) \log n - \xi(\boldsymbol{\gamma}^{(k)}, \boldsymbol{\gamma}^{(\ell)}). \tag{21}$$

Then (15), with $c_0 = \min\{c_j, j = 1, \ldots, p\}$, is the only form for a prior distribution that leads in a natural way to our approach being equivalent to a cost-adjusted version of BIC with the following properties: (a) if the cost of a variable $X_j$ is $\kappa$ times the minimum cost, then the imposed penalty $\omega$ is equivalent to adding $\kappa$ variables with the minimum cost; (b) our approach always results in models more parsimonious than BIC when costs are unequal, and (c) our prior reduces to BIC when all costs are equal (this result is summarized as Corollary 1 in Appendix A, where a proof is also provided).

**3. MCMC implementation.** With a realistically large number of predictors, the model space in our problem is too large for full-enumeration or naive Monte Carlo strategies to estimate posterior model probabilities with high accuracy in a reasonable amount of CPU time. For this reason, we adopted a different approach and implemented the following two-step method:

(1) First we used a model search tool to identify variables with high marginal posterior inclusion probabilities $f(\gamma_j|\mathbf{y})$, and we created a reduced model space consisting only of those variables whose marginal probabilities were above a threshold value. According to Barbieri and Berger (2004), this method of selecting variables based on their marginal probabilities may lead to the identification of models with better predictive abilities than approaches based on maximizing posterior model probabilities. Although Barbieri and Berger proposed 0.5 as a threshold value for $f(\gamma_j = 1|\mathbf{y})$, we used the lower value of 0.3, since our aim was only to identify and eliminate variables not contributing to models with high posterior probabilities.

(2) Then we used the same trans-dimensional MCMC algorithm as in step (1) in the reduced space to estimate posterior model probabilities (and the corresponding odds).

To ensure stability of our findings, we explored the use of two model search tools in step (1):

- a reversible-jump MCMC algorithm [RJMCMC; Green (1995)], as implemented for variable selection in generalized linear models by Dellaportas et al. (2002) and Ntzoufras et al. (2003); and
- the MCMC model composition ($MC^3$) algorithm [Madigan and York (1995)].

More specifically, we implemented reversible-jump moves within Gibbs for the model indicators $\gamma_j$, by proposing the new model to differ from the current one in each step by a single term $j$ with probability one [Dellaportas et al. (2002)]. Details on our RJMCMC and $MC^3$ implementations are given in Fouskakis, Ntzoutras and Draper (2009a).

The $MC^3$ approach relies on posterior model odds $PO_{\gamma,\gamma'}$, which are not analytically available in this setting; because of this we also explored two methods for calculating them—approximating the acceptance probabilities with cost-adjusted Laplace [equation (18)] and cost-adjusted BIC [equation (20)]—and, in addition, we further explored one additional form of sensitivity analysis: initializing the MCMC runs at the null model (with no predictors) and the full model (with all predictors). All of this was done both for the benefit-only analysis using our method (setting all costs equal) and the cost-benefit approach.

In moving from the full to the reduced model space to implement step (1) of our two-step method, for both the benefit-only and cost-benefit analyses we found a striking level of agreement, in the subset of variables defining the reduced model space, as we varied (a) the two model search tools, (b) the two methods to approximate the acceptance probabilities in $MC^3$, and (c) the two choices for initializing the MCMC runs; this made it unnecessary to perform similar sensitivity analyses in step (2). Results in the next section are therefore presented only for RJMCMC (starting from the full model). Convergence of the RJMCMC algorithm was checked using ergodic mean plots of the marginal inclusion probabilities for the full model space and the posterior model probabilities for the reduced space. Additional computing details are available in Appendix A.

In what follows we refer to the cost-benefit results as "RJMCMC cost-benefit," but we could equally well have used the term "$MC^3$ with cost-adjusted BIC" (or just "cost-adjusted BIC" for short), because the results from the two MCMC methods were in such close agreement.

TABLE 2
*Preliminary RJMCMC results: variables with marginal posterior probabilities $f(\gamma_j = 1|\mathbf{y})$ above 0.30; costs are expressed in minutes of abstraction time*

| | Variable | | Marginal posterior probabilities RJMCMC analysis | |
|---|---|---|---|---|
| Index | Name | Cost | Benefit-only | Cost-benefit |
| 1 | SBP score | 0.50 | 0.99 | 0.99 |
| 2 | Age | 0.50 | 0.99 | 0.99 |
| 3 | Blood urea nitrogen | 1.50 | 1.00 | 0.99 |
| 4 | APACHE II coma score | 2.50 | 1.00 | |
| 5 | Shortness of breath day 1? | 1.00 | 0.97 | 0.79 |
| 8 | Septic complications? | 3.00 | 0.88 | |
| 12 | Initial temperature | 0.50 | 0.98 | 0.96 |
| 13 | Heart rate day 1 | 0.50 | | 0.34 |
| 14 | Chest pain day 1? | 0.50 | | 0.39 |
| 15 | Cardiomegaly score | 1.50 | 0.71 | |
| 27 | Hematologic history score | 1.50 | 0.45 | |
| 37 | APACHE respiratory rate score | 1.00 | 0.95 | 0.32 |
| 46 | Admission SBP | 0.50 | 0.68 | 0.90 |
| 49 | Respiratory rate day 1 | 0.50 | | 0.81 |
| 51 | Confusion day 1? | 0.50 | | 0.95 |
| 70 | APACHE pH score | 1.00 | 0.98 | 0.98 |
| 73 | Morbid + comorbid score | 7.50 | 0.96 | |
| 78 | Musculoskeletal score | 1.00 | | 0.54 |

*Notes*: (1) Abbreviation used in this table: SBP = systolic blood pressure. (2) Variables with a question mark in their names were dichotomous answers to yes/no questions, scored 1 = yes and 0 = no; all other variables (except variable 1, which was also dichotomous) were measured on quantitative scales with three or more possible values.

## 4. Experimental results.

4.1. *Cost-benefit analysis with cost-adjusted BIC.* Table 2 presents the marginal posterior probabilities of the variables that exceeded the threshold value of 0.30, in each of the RJMCMC benefit-only and cost-benefit analyses in the reduced model space, together with their data collection costs. In both the benefit-only and cost-benefit settings our methods reduced the initial list of $p = 83$ available candidates down to 13 predictors. Note from Table 2 that the most expensive variables with high marginal posterior probabilities in the benefit-only analysis were absent from the set of promising variables in the cost-benefit analysis (e.g., the Morbid + comorbid score, variable 73). Similarly, some inexpensive variables with low marginal posterior probabilities in the benefit-only analysis were included in most of the models visited in the cost-benefit analysis (e.g., Confusion day 1?, variable 51). Note also that there is not a strong degree of overlap between the 14 variables cho-

TABLE 3
*Reduced model space: posterior model probabilities above 0.03, posterior odds ($PO_{1k}$) of the best model within each analysis versus the current model k, and model costs*

| k | Common variables within each analysis | Additional variables | | | | | Model cost | Posterior probabilities | $PO_{1k}$ |
|---|---|---|---|---|---|---|---|---|---|
| Benefit-only analysis | | | | | | | | | |
| 1 | $X_4 + X_{15} + X_{37} + X_{73}$ | $+X_8$ | $+X_{27}$ | $+X_{46}$ | | | 22.5 | 0.3066 | 1.00 |
| 2 | | $+X_8$ | $+X_{27}$ | | | | 22.0 | 0.1969 | 1.56 |
| 3 | | $+X_8$ | | | | | 20.5 | 0.1833 | 1.67 |
| 4 | | | $+X_{27}$ | $+X_{46}$ | | | 19.5 | 0.0763 | 4.02 |
| 5 | | | | | | | 17.5 | 0.0383 | 8.00 |
| Cost-benefit analysis | | | | | | | | | |
| 1 | $X_{46} + X_{51}$ | | | | $+X_{49}$ | $+X_{78}$ | 7.5 | 0.1460 | 1.00 |
| 2 | | | $+X_{14}$ | | $+X_{49}$ | $+X_{78}$ | 7.5 | 0.1168 | 1.27 |
| 3 | | $+X_{13}$ | | | $+X_{49}$ | $+X_{78}$ | 7.5 | 0.0866 | 1.69 |
| 4 | | $+X_{13}$ | $+X_{14}$ | | $+X_{49}$ | $+X_{78}$ | 8.0 | 0.0665 | 2.20 |
| 5 | | | $+X_{14}$ | | $+X_{49}$ | | 7.0 | 0.0461 | 3.17 |
| 6 | | | | | $+X_{49}$ | | 6.5 | 0.0409 | 3.57 |
| 7 | | | | $+X_{37}$ | | $+X_{78}$ | 7.5 | 0.0382 | 3.82 |
| 8 | | $+X_{13}$ | $+X_{14}$ | | $+X_{49}$ | | 7.5 | 0.0369 | 3.96 |
| 9 | | $+X_{13}$ | | | | | 6.5 | 0.0344 | 4.25 |

Common variables in both analyses: $X_1 + X_2 + X_3 + X_5 + X_{12} + X_{70}$.

sen in the original RAND benefit-only analysis summarized in Table 1 and the 13 variables with high marginal posterior probabilities in the RJMCMC benefit-only part of Table 2; we return to this point below.

Table 3 presents models with posterior model probabilities above 0.03 (in descending order), as well as posterior odds of the model with the highest posterior probability compared to the remaining ones. In both types of analysis, the variables Systolic blood pressure score ($X_1$), Age ($X_2$), Blood urea nitrogen ($X_3$), Shortness of breath ($X_5$), Temperature ($X_{12}$) and APACHE pH score ($X_{70}$) were included in all the highest probability models, with costs (in minutes) 0.5, 0.5, 1.5, 1.0, 0.5 and 1.0 respectively.

For the cost-benefit analysis, 9 models had posterior probabilities above 0.03. In all of these models Admission systolic blood pressure (SBP; $X_{46}$) and Confusion day 1 ($X_{51}$) were present (both having the lowest cost of 0.5 minutes). Predictors Respiratory rate day 1 ($X_{49}$) and Musculoskeletal score ($X_{78}$) were also frequently included in the top nine models (in 7 and 5 of the 9 cases, respectively). Both of these variables were present in the four highest probability models, with similar posterior probabilities; in fact, there were no substantial differences between those models (note that the posterior odds of models 2–4 in the cost-benefit part of Table 3 differed from those of model 1 by factors of no more than three, indicating evidence "not worth more than a bare mention" [cf. Raftery (1996)] in favor of model 1). All variables included in the highest probability models had costs of at most one minute with the exception of Blood urea nitrogen ($X_3$), which had a cost of 1.5.

In the RJMCMC benefit-only analysis, 5 models had posterior probabilities above 0.03. In all of these models APACHE II coma score ($X_4$), Cardiomegaly score ($X_{15}$), APACHE respiratory rate score ($X_{37}$) and Morbid + comorbid score ($X_{73}$) were present, having costs of 2.5, 1.5, 1.0 and 7.5 minutes (respectively). Note that the costs of the best models in the benefit-only analysis are 2.2 to 3.5 times higher than the costs of the best models from the cost-benefit analysis.

Since in the RJMCMC cost-benefit analysis we increase the penalty of relatively expensive variables in the prior, we end up selecting more parsimonious models in terms of both dimensionality and cost. It is therefore interesting to examine the loss in terms of prediction and goodness of fit. We use the posterior distribution of the deviance statistic

$$D(\boldsymbol{\beta}_{\boldsymbol{\gamma}}, \boldsymbol{\gamma}) = -2 \sum_{i=1}^{n} \log f(y_i | \boldsymbol{\beta}_{\boldsymbol{\gamma}}, \boldsymbol{\gamma}) \tag{22}$$

[Dempster (1974); Spiegelhalter et al. (2002)] as a measure of model fit. Usually attention focuses on the minimum value of this posterior distribution (which sometimes is poorly estimated by MCMC runs), but other posterior descriptive measures such as the median or mean provide adequate measures of fit [Spiegelhalter et al. (1996)].

In Table 4 we present the minimum and median values of the posterior distribution of the deviance statistic, together with the cost and dimension of the best models found with both types of analysis. Two main points are worth noting:

- Two approaches to the creation of a benefit-only analysis may now be compared: the frequentist approach employed in the original RAND study and our RJMCMC benefit-only analysis obtained by setting all the costs equal. The deviance statistic for the benefit-only RAND model summarized in Table 1 turned out to be 1587.3 (achieved with 14 predictors), substantially worse than the median deviance (1564.5, achieved with 13 predictors) of the best model visited by the benefit-only approach examined in this paper; in other words, in this case study, frequentist backward selection from the model with all predictors (the RAND approach) was substantially out-performed by Bayesian RJMCMC.
- The minimum and median values of the posterior distribution of the deviance statistic for the RJMCMC benefit-only analysis were lower by a relatively modest 5.3% and 5.1% compared to the corresponding values of the cost-benefit analysis, but the cost of the best model in the cost-benefit

Table 4

*Comparison of measures of fit, cost and dimensionality between the best models in the reduced model space of the RJMCMC benefit-only and cost-benefit analyses; percentage difference is in relation to benefit-only*

|  | RJMCMC analysis | | Percentage difference |
|---|---|---|---|
|  | **Benefit-only** | **Cost-benefit** |  |
| Minimum deviance | 1553.2 | 1635.8 | $+5.3$ |
| Median deviance | 1564.5 | 1644.8 | $+5.1$ |
| Cost | 22.5 | 7.5 | $-66.7$ |
| Dimension | 13 | 10 | $-23.1$ |

analysis was almost 67% lower than that for the benefit-only analysis; similarly, the dimensionality of the best model in the cost-benefit analysis was about 23% lower than that for the benefit-only analysis. These values indicate that the loss of predictive accuracy with the cost-benefit analysis is small compared to the substantial gains achieved in cost and reduced model complexity.

An alternative predictive measure of fit is the cross-validation log score $LS_{CV}$, following ideas of Geisser and Eddy (1979) and Gelfand et al. (1992) [also see, e.g., Draper and Krnjajić (2009)]. It is based on leave-one-out predictive distributions $f(y_i|\mathbf{y}_{\setminus i})$ and is given by

$$LS_{CV}(\boldsymbol{\gamma}|\mathbf{y}) = \frac{1}{n}\sum_{i=1}^{n} \log f(y_i|\mathbf{y}_{\setminus i}, \boldsymbol{\gamma}), \qquad (23)$$

where $\mathbf{y}_{\setminus i}$ is the vector of data $\mathbf{y}$ without observation $i$ (larger values of $LS_{CV}$ indicate greater predictive accuracy). This measure can be estimated directly from a single MCMC run using the formula

$$\widehat{LS}_{CV}(\boldsymbol{\gamma}|\mathbf{y}) = -\frac{1}{n}\sum_{i=1}^{n}\log \overline{f^{-1}(y_i|\boldsymbol{\beta}_{\boldsymbol{\gamma}}, \boldsymbol{\gamma})}, \qquad (24)$$

where $\overline{f^{-1}(y_i|\boldsymbol{\beta}_{\boldsymbol{\gamma}}, \boldsymbol{\gamma})}$ is the posterior mean of the inverse of the predictive density for observation $i$ [for details see, e.g., Gelfand (1996), pages 154–155]. We calculated $\widehat{LS}_{CV}$ for the models with the highest posterior probability for each RJMCMC analysis and obtained a value of $-0.312$ for the best model of the benefit-only analysis and $-0.327$ for that of the cost-benefit analysis; the latter is 4.8% smaller than the former, in line with the last column of Table 4, and, as before, this small loss in predictive accuracy is accompanied by the 66% drop in cost and 15% decrease in model complexity achieved by the cost-benefit approach.

TABLE 5
*The full set of 83 variables, together with their data collection costs per patient and their status according to the utility and RJMCMC methods; columns 4 and 5 are explained in the text*

| | Variable | | Method | |
|---|---|---|---|---|
| Index | Name | Cost $c_j$ (minutes) | Utility | RJMCMC |
| 1 | Systolic BP score | 0.5 | ** | ** |
| 2 | Age | 0.5 | | ** |
| 3 | Blood urea nitrogen | 1.5 | ** | ** |
| 4 | APACHE II coma score | 2.5 | ** | |
| 5 | Shortness of breath day 1? | 1.0 | * | ** |
| 6 | Serum albumin score | 1.5 | | |
| 7 | Respiratory distress? | 1.0 | | |
| 8 | Septic complications? | 3.0 | | |
| 9 | Prior respiratory failure? | 2.0 | | |
| 10 | Recently hospitalized? | 2.0 | | |
| 11 | Racbilateral process score | 1.5 | | |
| 12 | Initial temperature | 0.5 | ** | ** |
| 13 | Heart rate day 1 | 0.5 | * | * |
| 14 | Chest pain day 1? | 0.5 | * | * |
| 15 | Cardiomegaly score | 1.5 | | |
| 16 | Plural effusion score | 1.5 | | |
| 17 | CXR CHF score | 2.5 | | |
| 18 | Ambulatory score | 2.5 | | |
| 19 | Endocarditis at admission? | 1.5 | | |
| 20 | CPK score | 2.0 | | |
| 21 | Prior antibiotics? | 0.5 | | |
| 22 | Prior interstitial lung disease? | 0.5 | | |
| 23 | Home oxygen use? | 1.0 | | |
| 24 | Prior pneumonectomy? | 0.5 | | |
| 25 | Prior tracheostomy? | 0.5 | | |
| 26 | Prior aminophylline score | 0.5 | | |
| 27 | Hematologic history score | 1.5 | | |
| 28 | Cancer score | 1.5 | | |
| 29 | APACHE heart rate score | 1.5 | | |
| 30 | Corodaker score | 1.0 | | |
| 31 | Disease of thorax? | 1.0 | | |
| 32 | Multiple myeloma? | 0.5 | * | |
| 33 | Immunocompromised? | 0.5 | | |
| 34 | Residence score | 1.0 | | |
| 35 | Hepatobiliary history? | 0.5 | | |
| 36 | Renal history score | 1.0 | | |
| 37 | APACHE respiratory rate score | 1.0 | * | * |
| 38 | New lung score | 1.0 | | |
| 39 | Co-morbid aspiration score | 0.5 | * | |

TABLE 5
*(Continued.)*

| | Variable | | Method | |
|---|---|---|---|---|
| Index | Name | Cost $c_j$ (minutes) | Utility | RJMCMC |
| 40 | APACHE sodium score | 2.0 | | |
| 41 | APACHE hematocrit score | 1.5 | | |
| 42 | APACHE WBC score | 1.5 | | |
| 43 | APACHE oxygenation score | 1.5 | | |
| 44 | CVA score | 1.0 | | |
| 45 | APACHE potassium score | 1.0 | | |
| 46 | Admission SBP | 0.5 | ** | ** |
| 47 | CHF Chest X-ray score | 2.5 | | |
| 48 | Total APACHE II score | 10.0 | | |
| 49 | Respiratory rate day 1 | 0.5 | ** | ** |
| 50 | DBP day 1 | 0.5 | | |
| 51 | Confusion day 1? | 0.5 | * | ** |
| 52 | PVC score | 0.5 | | |
| 53 | APACHE VB score | 1.5 | | |
| 54 | Pulmonary edema score | 0.5 | | |
| 55 | Sum of CHF components | 5.5 | | |
| 56 | Influenza score | 0.5 | * | |
| 57 | Arrest in ER score | 0.5 | * | |
| 58 | Biliribin score | 1.5 | | |
| 59 | Positive blood culture? | 0.5 | | |
| 60 | Positive urine culture? | 0.5 | | |
| 61 | Wheezing at admission? | 0.5 | | |
| 62 | Body system count | 2.5 | | |
| 63 | Morbid prior COPD score | 0.5 | | |
| 64 | Morbid PHS | 0.5 | | |
| 65 | Co-morbid cirrhosis score | 0.5 | * | |
| 66 | Co-morbid CHF score | 0.5 | * | |
| 67 | Co-morbid arrhythmias score | 0.5 | | |
| 68 | Co-morbid smoking score | 0.5 | | |
| 69 | Co-morbid alcoholism score | 0.5 | * | |
| 70 | APACHE pH score | 1.0 | | ** |
| 71 | Co-morbid NGTs score | 0.5 | | |
| 72 | Co-morbid steroids score | 0.5 | * | |
| 73 | Morbid + comorbid score | 7.5 | | |
| 74 | Cardiac history score | 0.5 | | |
| 75 | Neurologic history score | 0.5 | * | |
| 76 | Oncologic history score | 0.5 | | |
| 77 | Immunologic history score | 0.5 | | |
| 78 | Musculoskeletal score | 0.5 | * | ** |

TABLE 5
*(Continued.)*

| | Variable | | Method | |
|---|---|---|---|---|
| Index | Name | Cost $c_j$ (minutes) | Utility | RJMCMC |
| 79 | APACHE temperature score | 1.0 | | |
| 80 | APACHE mean BP score | 1.0 | | |
| 81 | APACHE creatinine score | 1.0 | | |
| 82 | DX score | 1.0 | | |
| 83 | Sex of patient | 0.5 | | |

*Notes:* (1) Abbreviations used in this table: BP = blood pressure, CHF = congestive heart failure, COPD = chronic obstructive pulmonary disease, CVA = cardiovascular accident, CXR = chest X-ray, DBP = diastolic blood pressure, DX = diagnoses, NGT = naso-gastric tube, PHS = pulmonary hospitalization score, PVC = pulmonary vascular congestion, SBP = systolic blood pressure, VB = venous bicarbonate. (2) Variables with a question mark in their names were dichotomous answers to yes/no questions, scored 1 = yes and 0 = no; all other variables (except variable 1, which was also dichotomous) were measured on quantitative scales with three or more possible values.

4.2. *Comparison of cost-benefit analyses: utility versus cost-adjusted BIC.* The RJMCMC cost-benefit results so far have been based on our cost-adjusted BIC approach; other cost-benefit methods are possible, including a decision-theoretic approach involving (a) explicit quantification of costs and benefits via a utility function followed by (b) maximization of expected utility. Columns 4–6 in Table 1 present a comparison of the maximum-expected-utility method of Fouskakis and Draper (2008)[3] and the RJMCMC method of this paper—which (as noted above) is functionally equivalent to our cost-adjusted BIC method—when the space of predictor variables is defined by the $p = 14$ variables in the original RAND scale described in Section 1, and Table 5 offers a comparison of the two methods when all $p = 83$ variables collected in the RAND study served as the basis of the variable selection search. In columns 4 and 5 of Table 1, two stars signify that a variable appeared in the best model found by each method [for RJMCMC this model is given in Table 6(B)], and one star in column 4 means that the variable often appeared in the 20 best models identified by the utility approach.

It is evident that the two methods arrived at similar conclusions with $p = 14$: six variables were chosen by neither method (note from column 6

---

[3]Fouskakis and Draper (2008) used stochastic optimization methods—including simulated annealing, genetic algorithms and tabu search [see Draper and Fouskakis (2000), and Fouskakis and Draper (2002)]—to find (near-) optimal subsets of predictor variables that maximize an expected utility function that trades off data collection cost against predictive accuracy in a way that is sensitive to the policy implications of searching for "good" and "bad" hospitals; utility elicitation details are available in Fouskakis, Ntzoufras and Draper (2009a).

Table 6

(A) *Summary of the RJMCMC cost-benefit search results in the $p = 14$ case.* (B) *Comparison of the utility and RJMCMC methods in how their best models trade off cost and predictive accuracy*

A

| Model | Cost | Posterior probability | $PO_{1k}$ |
|---|---|---|---|
| $X_1 + X_2 + X_3 + X_4 + X_5 + X_6 + X_7 + X_{12}$ | 9.0 | 0.453 | 1.00 |
| $X_1 + X_2 + X_3 + X_4 + X_5 \phantom{+ X_6} + X_7 + X_{12}$ | 7.5 | 0.415 | 1.09 |
| $X_1 + X_2 + X_3 + X_4 + X_5 + X_6 \phantom{+ X_7} + X_{12}$ | 8.0 | 0.054 | 8.40 |
| $X_1 + X_2 + X_3 + X_4 + X_5 + X_6 + X_7$ | 8.5 | 0.031 | 14.72 |

B

| $p$ | Method | Model | Cost | Median deviance | $LS_{CV}$ |
|---|---|---|---|---|---|
| 14 | RJMCMC | $X_1 + X_2 + X_3 + X_4 + X_5 + X_6 + X_7 + X_{12}$ | 9.0 | 1654 | $-0.329$ |
|  |  | $X_1 + X_2 + X_3 + X_4 + X_5 \phantom{+ X_6} + X_7 + X_{12}$ | 7.5 | 1676 | $-0.333$ |
|  | Utility | $X_1 \phantom{+ X_2} + X_3 + X_4 + X_5$ | 5.5 | 1726 | $-0.342$ |
| 83 | RJMCMC | $X_1 + X_2 + X_3 \phantom{+ X_4} + X_5 \phantom{+ X_6 + X_7} + X_{12}$ $+ X_{46} + X_{49} + X_{51} \phantom{+ X_{57}} + X_{70} + X_{78}$ | 7.5 | 1645 | $-0.327$ |
|  | Utility | $X_1 \phantom{+ X_2} + X_3 + X_4 \phantom{+ X_5 + X_6 + X_7} + X_{12}$ $+ X_{46} + X_{49} \phantom{+ X_{51}} + X_{57}$ | $6.5^*$ | 1693 | $-0.336$ |

$^*$This model had a higher cost than the best model with $p = 14$ because the utility approach was not optimizing on cost but on a utility-based cost-benefit tradeoff.

in Table 1 that all of these variables had marginal posterior probability 0 in the RJMCMC method); four more variables had identical star patterns; and the other four variables were chosen by both methods as important, differing only in how many stars they received. Table 5 uses a similar star system: two stars in columns 4 and 5 in this table signify membership in the globally best model found by the utility and RJMCMC methods, respectively; one star in column 4 means that the variable appeared frequently in the 100 best utility models [see Fouskakis and Draper (2008) for details], with one star in column 5 signifying that a variable often occurred in the highest-posterior-probability RJMCMC models of Table 3. With $p = 83$, the agreement between the two methods is also strong (although not as strong as with $p = 14$): 60 variables were ignored by both methods, eight variables had identical star patterns, three variables were chosen as important by both methods but with different star patterns, 10 variables were marked as important by the utility approach and not by RJMCMC, and two variables were singled out by RJMCMC and not by utility (this represents substantial agreement on the importance of 85% of the variables).

Table 6 gives a summary of the RJMCMC search results with $p = 14$ and examines the cost-benefit tradeoffs of the utility and RJMCMC methods in more detail. It is clear that, to the extent that the two approaches differ, the utility method favors models that cost somewhat less but also predict somewhat less well. The fact that the two methods yield somewhat different results does not mean that either is wrong; they are both valid solutions to similar but not identical problems. Both methods lead to noticeably better models (in a cost-benefit sense) than frequentist or Bayesian benefit-only approaches, when—as is often the case—cost is an issue that must be included in the problem formulation to arrive at a policy-relevant solution.

**5. Discussion.** In this paper we have examined a relatively new perspective on Bayesian variable selection, when data collection costs need to be traded off against predictive accuracy in choosing an optimal subset of predictors. We propose a prior setup that accounts for the cost of each variable and we utilize traditional posterior model odds for the evaluation of models. This leads to a set of posterior model probabilities that correspond approximately to a generalized cost-adjusted version of BIC. Computation is performed using reversible-jump MCMC in two stages: first, to reduce the model space by dropping variables with low marginal posterior probabilities, and then to estimate posterior model probabilities in the reduced space. We have applied our methodology to the problem of cost-effective input-output quality measurement in a health policy setting, with a binary outcome and a large number ($p = 83$) of predictors that differ substantially in data-collection costs. The resulting models achieve dramatic gains in cost and noticeable improvement in model simplicity at the price of a small loss in predictive accuracy, when compared to the results of a more traditional benefit-only analysis.

As noted in Section 1.3, the problem we address here can also be approached via an alternative version of cost-benefit analysis (based on maximizing expected utility) and/or a cost-restriction-benefit analysis (maximizing predictive accuracy subject to a bound on cost); the latter approach [Fouskakis, Ntzoufras and Draper (2009b)] is not directly comparable to the other two methods. We have compared our cost-adjusted BIC approach with the utility-based method developed by Fouskakis and Draper (FD) (2008), finding that the two approaches lead to similar cost-benefit variable subsets. The decision-theoretic approach has the drawback that it may not be possible to find a single utility structure capturing the preferences of all relevant stakeholders (including patients, doctors, hospitals, citizen watchdog groups, and state and federal regulatory agencies) in the quality-of-care-assessment process (utility was assessed from only a single viewpoint in FD). The cost-adjusted BIC approach developed here offers an alternative that avoids ambiguity in utility specification.

The method we have described in this paper appears to hold significant promise for cost-effective input-output quality and performance assessment.

It can be applied in any setting where the outcome is binary, such as in education (with outcomes such as drop-out during university study and employment following graduation) and business (with outcomes such as retention in the workplace and the default status of a loan), and can be implemented with minor modifications for any other generalized linear model. We believe that the scope of applications of regression methodology in which

(a) the purpose of the model-building is to create a predictive scale and
(b) future use of the scale created in (a) will take place in a cost-constrained environment with nonzero data collection costs

is sufficiently broad that methods like those examined here are worthy, both of consideration now, for practical adoption, and of further study to promote, for example, additional computational efficiency gains.

## APPENDIX A

Here we give a proof of Theorem 1 and the statement and proof of Corollary 1.

PROOF OF THEOREM 1. To begin with condition (a) in Section 2.2.2, it is straightforward to show that this is satisfied if and only if the costs enter into the prior through functions of the ratios $\frac{c_j}{c_{j'}}$, or equivalently, through functions of ratios $\frac{c_j}{c_0}$ for some $c_0 > 0$.

From condition (b), the extra penalty $\xi_1$ when comparing two models $(\gamma_j = 1, \boldsymbol{\gamma}_{\setminus j})$ and $(\gamma_j = 0, \boldsymbol{\gamma}_{\setminus j})$ that differ only by variable $X_j$ with cost $c_0$ is given by

$$\xi_1[(\gamma_j = 1, \boldsymbol{\gamma}_{\setminus j}), (\gamma_j = 0, \boldsymbol{\gamma}_{\setminus j})] = -2 \log \frac{f(\gamma_j = 1, \boldsymbol{\gamma}_{\setminus j})}{f(\gamma_j = 0, \boldsymbol{\gamma}_{\setminus j})} = 0$$

(25)
$$\Leftrightarrow \quad \frac{f(\gamma_j = 1 | \boldsymbol{\gamma}_{\setminus j})}{1 - f(\gamma_j = 1 | \boldsymbol{\gamma}_{\setminus j})} = 1 \quad \Leftrightarrow \quad f(\gamma_j = 1 | \boldsymbol{\gamma}_{\setminus j}) = \frac{1}{2}.$$

Since the above must be true for all $\gamma_j$ and $\boldsymbol{\gamma}_{\setminus j}$, the result is $f(\gamma_j = 1) = \frac{1}{2}$ when the cost of a variable $X_j$ equals the baseline cost $c_0$.

Similarly, for two models $(\gamma_j = 1, \boldsymbol{\gamma}_{\setminus j})$ and $(\gamma_j = 0, \boldsymbol{\gamma}_{\setminus j})$ that differ only by variable $X_j$ with cost $c_j = \kappa c_0$, from condition (c) we have

$$\xi_2[(\gamma_j = 1, \boldsymbol{\gamma}_{\setminus j}), (\gamma_j = 0, \boldsymbol{\gamma}_{\setminus j})] = -2 \log \frac{f(\gamma_j = 1 | \boldsymbol{\gamma}_{\setminus j})}{f(\gamma_j = 0 | \boldsymbol{\gamma}_{\setminus j})} = (\kappa - 1) \log n$$

$$\Leftrightarrow \quad f(\gamma_j = 1 | \boldsymbol{\gamma}_{\setminus j}) = \frac{\exp[-(1/2)(\kappa - 1) \log n]}{1 + \exp[-(1/2)(\kappa - 1) \log n]}.$$

Since the above equation must hold for any $\gamma_j, \boldsymbol{\gamma}_{\setminus j}$ and $\kappa = \frac{c_j}{c_0}$, we end up with a prior of the form (15).

From condition (d) we have to set $c_0 \leq \min\{c_j, j = 1, \ldots, p\}$, since any other choice will result in negative prior penalties for variables with cost less than $c_0$. Furthermore, if $c_j = c' \geq c_0$ for all $j$, then from (c),

$$f(\gamma_j = 1) = \frac{\exp[-(1/2)(c'/c_0 - 1)\log n]}{1 + \exp[-(1/2)(c'/c_0 - 1)\log n]}, \tag{26}$$

resulting in

$$f(\boldsymbol{\gamma}) \propto \frac{\{\exp[-(1/2)(c'/c_0 - 1)\log n]\}^{\sum_{j=1}^p \gamma_j}}{\{1 + \exp[-(1/2)(c'/c_0 - 1)\log n]\}^p}. \tag{27}$$

Under condition (e), for any two models $\boldsymbol{\gamma}^{(k)}$ and $\boldsymbol{\gamma}^{(\ell)}$ the prior model odds must equal to one. Hence,

$$\frac{f(\boldsymbol{\gamma}^{(k)})}{f(\boldsymbol{\gamma}^{(\ell)})} = 1, \tag{28}$$

resulting in

$$\sum_{j=1}^p (\gamma_j^{(k)} - \gamma_j^{(\ell)})\left[-\frac{1}{2}\left(\frac{c'}{c_0} - 1\right)\log n\right] = 0 \qquad \text{for any } \boldsymbol{\gamma}^{(k)}, \boldsymbol{\gamma}^{(\ell)}. \tag{29}$$

The above is satisfied for any pair of models if and only if $c' = c_0$. Hence, $c_0 = \min\{c_j, j = 1, \ldots, p\}$ is the only choice under which (e) is satisfied when (b), (c) and (d) also hold. $\square$

COROLLARY 1. *If a prior distribution $f(\gamma)$ is such that:*

(a′) *the imposed penalty $\omega$ on the log-likelihood ratio for adding a variable $X_j$ with cost $\kappa$ times the baseline cost $c_0$ (for positive integer $\kappa$) equals the imposed penalty for adding $\kappa$ variables with the baseline cost $c_0$,*
(b′) *the imposed penalty $\omega$ for each additional variable is at least equal to $\log n$ (the BIC penalty for the benefit-only analysis), and*
(c′) *if all costs are equal, the imposed penalty $\omega$ when comparing any two models $\boldsymbol{\gamma}^{(k)}$ and $\boldsymbol{\gamma}^{(\ell)}$ is $(d_{\boldsymbol{\gamma}^{(k)}} - d_{\boldsymbol{\gamma}^{(\ell)}})\log n$ (the BIC penalty),*

*then it must be of the form (15).*

PROOF. If we compare two models that differ only by a variable $X_j$ with cost $c_j = \kappa c_0$, then from (a′),

$$\omega[(\gamma_j = 1, \boldsymbol{\gamma}_{\setminus j}), (\gamma_j = 0, \boldsymbol{\gamma}_{\setminus j})] = \kappa \log n \qquad \text{for all } j = 1, \ldots, p, \tag{30}$$

which results in

$$\xi[(\gamma_j = 1, \boldsymbol{\gamma}_{\setminus j}), (\gamma_j = 0, \boldsymbol{\gamma}_{\setminus j})] = (\kappa - 1)\log n \qquad \text{for all } j = 1, \ldots, p, \tag{31}$$

from (21). The above expression corresponds to the third requirement of Theorem 1 used to construct our proposed prior distribution. Moreover, the second requirement of the same theorem arises as a special case for $\kappa = 1$.

From (b′) we have that

$$\text{(32)} \quad \omega[(\gamma_j = 1, \boldsymbol{\gamma}_{\setminus j}), (\gamma_j = 0, \boldsymbol{\gamma}_{\setminus j})] \geq \log n \quad \text{for all } j = 1, \ldots, p,$$

resulting in

$$\text{(33)} \quad \xi[(\gamma_j = 1, \boldsymbol{\gamma}_{\setminus j}), (\gamma_j = 0, \boldsymbol{\gamma}_{\setminus j})] \geq 0 \quad \text{for all } j = 1, \ldots, p,$$

which corresponds to the fourth requirement of Theorem 1.

Finally, from (c′), if all costs are equal, then

$$\text{(34)} \quad \omega(\boldsymbol{\gamma}^{(k)}, \boldsymbol{\gamma}^{(\ell)}) = (d_{\boldsymbol{\gamma}^{(k)}} - d_{\boldsymbol{\gamma}^{(\ell)}}) \log n,$$

resulting in

$$\text{(35)} \quad \xi(\boldsymbol{\gamma}^{(k)}, \boldsymbol{\gamma}^{(\ell)}) = 0 \quad \text{for all } \boldsymbol{\gamma}^{(k)} \text{ and } \boldsymbol{\gamma}^{(\ell)},$$

and thus,

$$\text{(36)} \quad \frac{f(\boldsymbol{\gamma}^{(k)})}{f(\boldsymbol{\gamma}^{(\ell)})} = 1$$

for any pair of compared models. Hence, the induced prior when all costs are equal must be uniform on $\boldsymbol{\gamma}$, that is, $f(\boldsymbol{\gamma}) \propto 1$ for all $\boldsymbol{\gamma}$. This corresponds to the fifth requirement used to construct our prior in Theorem 1.

Since each of the above statements is equivalent to the requirements used to build our prior in Theorem 1, the only prior with the above three properties is (15). □

## APPENDIX B: COMPUTING DETAILS FOR THE MCMC-BASED COST-BENEFIT ANALYSIS

With reference to the MCMC methods described in Section 3, both the coding time and the running time of RJMCMC were higher than with either variant of $MC^3$ to achieve comparable MCMC accuracy. All $MC^3$ runs in the full model space were based on 10,000 monitoring iterations after a burn-in (from either the null model or the full model) of 1,000 iterations; each of these runs took 2–3 days (on a Pentium 4 machine at 2.8 GHz with 512 MB RAM) for the cost-adjusted Laplace variant of $MC^3$ and 1–2 days for the cost-adjusted BIC variant (these are run times for an implementation in R; coding the same algorithms in C would have yielded substantially faster run times, on the order of 6–8 hours for Laplace and 2–5 hours for BIC). To achieve reasonable run times for RJMCMC, it was necessary to implement the algorithm in C. RJMCMC runs were based on 100,000 iterations, after discarding an initial 10,000 iterations as a burn-in; each of these runs took 2–3 days in the full model space and 9 hours in the reduced space. The resulting R and C programs are available upon request from the first or second authors of this paper.

**Acknowledgments.** The authors are grateful to Katherine Kahn for providing them with data from the RAND DRG Quality of Care study, and to the Editor and referees for comments that substantially improved the paper.

## SUPPLEMENTARY MATERIAL

**Cost-based prior distributions for variable selection in generalized linear models** (DOI: [10.1214/08-AOAS207SUPP](10.1214/08-AOAS207SUPP); .pdf). Imaginary data and power-prior motivation for the prior distribution in the main paper's equation (6) and details on RJMCMC and $MC^3$ implementation and utility elicitation.

D. Fouskakis  
Department of Mathematics  
National Technical University of Athens  
Zografou Campus, Athens 15780  
Greece  
E-mail: [fouskakis@math.ntua.gr](fouskakis@math.ntua.gr)

I. Ntzoufras  
Department of Statistics  
Athens University of Economics  
and Business  
76 Patision Street, Athens 10434  
Greece  
E-mail: [ntzoufras@aueb.gr](ntzoufras@aueb.gr)

D. Draper
Department of Applied Mathematics
    and Statistics
Baskin School of Engineering
University of California
1156 High Street, Santa Cruz,
    California 95064
USA
E-mail: draper@ams.ucsc.edu